\documentclass[%
amsmath,amssymb,
aps,
prfluids,
]{revtex4-2}

\usepackage{xcolor}
\usepackage{soul}
\usepackage{bm}
\usepackage[normalem]{ulem}
\usepackage{graphicx}
\usepackage{dcolumn}
\usepackage{xcolor}
\usepackage{mathptmx}
\usepackage{etoolbox}
\usepackage{appendix}

\usepackage{multirow}

\begin{document}

\preprint{APS/123-QED}

\title{Toroidal cavitation by a snapping popper }

\author{Akihito Kiyama}
\author{Sharon Wang}
\author{Sunghwan Jung}
\affiliation{Department of Biological and Environmental Engineering, Cornell University, Ithaca, New York 14850, USA}
\email{sj737@cornell.edu}


\date{\today}
\begin{abstract}
Cavitation is a phenomenon in which bubbles form and collapse in liquids due to pressure or temperature changes. Even common tools like a rubber popper can be used to create cavitation at home. As a rubber popper toy slams a solid wall underwater, toroidal cavitation forms. As part of this project, we aim to explain how an elastic shell causes cavitation and to describe the bubble morphology. High-speed imaging reveals that a fast fluid flow between a snapping popper and a solid glass reduces the fluid pressure to cavitate. Cavitation occurs on the popper surface in the form of sheet cavitation. Our study uses two-dimensional Rayleigh-Plesset equations and the energy balance to capture the relationship between the bubble lifetime and the popper deformability. The initial distance between the popper and the wall is an important parameter for determining the cavitation dynamics. Presented results provide a deeper understanding of cavitation mechanics, which involves the interaction between fluid and elastic structure.

\end{abstract}

\maketitle

\section{Introduction}
Cavitation is a phase change process from liquid to gas (i.e., vaporization) due to an abrupt decrease in fluid pressure. Cavitation bubbles collapse onto solid objects and walls after nucleation, causing destructive erosion. For example, cavitation in a high-speed flow around a hydrofoil or a propeller damages their structure (e.g., \cite{Brennen1995,Teran2018,Teran2019}). It can significantly reduce the efficiency of marine propulsion and hydro turbine systems and cause design failures due to the excessive vibration.  On the other hand, engineers use the impulsive fluid motion associated with cavitation bubbles in beneficial ways, e.g. for medical \cite{Maxwell2011} and cleaning applications \cite{Ohl2006,Lee2018}. 

Researchers have been employing various experimental methods to create a spherical cavitation bubble \cite{Blake1987}. For example, a short-pulsed laser was used to study bubble collapse and rebound behaviors \cite{Supponen2016,Supponen2017,Supponen2018}. The laser method allows researchers to study the high-precision behavior of cavitation bubbles down to nanosecond time scales. The electric spark method is used to create a single bubble at a low cost, as used in the study for bubble-particle interaction \cite{Poulain2015} and bubble dynamics in non-Newtonian fluid \cite{Bokman2022}. Ultrasonic transducers have also been used as an alternative to electric spark method to create bubble cavitation \cite{Znidarcic2014,Morton2023}. 

Cavitation occurs not only in engineering systems but also in natural systems and everyday items. In nature, pistol shrimp use bubbles to stun their prey \cite{Versluis2000,Koukouvinis2017}. These bubbles are created by the shrimp's claw and can reach high temperatures and high pressure, killing small fish. Even in everyday activities, cavitation can be seen when one cracks the finger joint \cite{Kawchuk2015}, drops a water-filled vials \cite{Randolph2015}/tubes \cite{Kiyama2016}, or performs a party trick with a beer bottle \cite{Pan2017}.  Such events are caused by the formation of bubbles and their subsequent collapse, which release energy in the form of shockwaves with loud sound and heat.

\begin{figure*}
    \centering
\includegraphics[width=0.75\paperwidth]{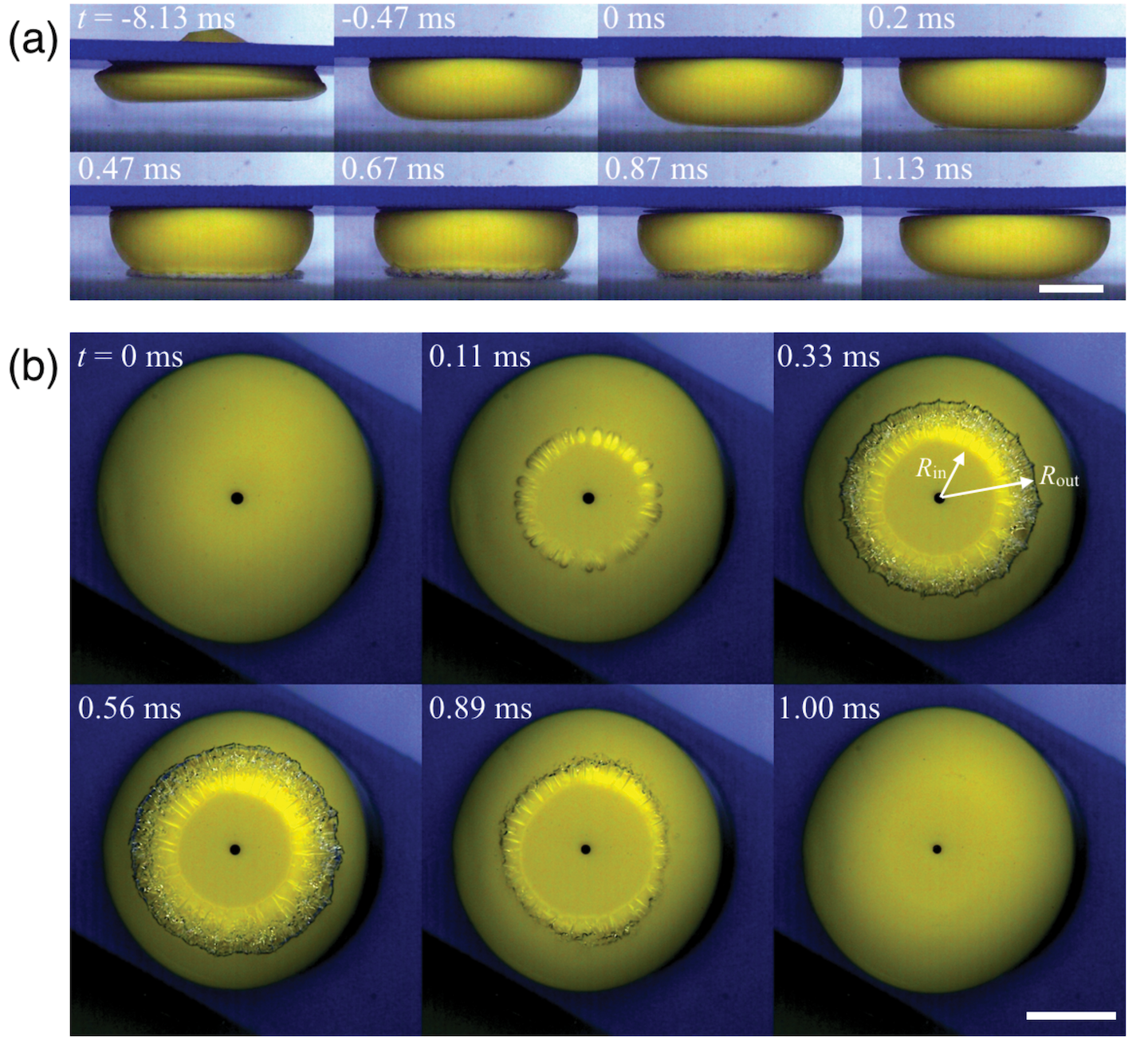}
    \caption{(a) side-view images of the underwater popper ($R_p\approx16$~mm) approaching the glass substrate. The platform height was set at $H\approx9$~mm (i.e., $H/R_p\approx0.56$). (b) bottom-view images of the same popper and cavitation dynamics. We used 50\% glycerol-water mixture ($\approx 5$~cSt) for visualization purpose. The platform height was set at $H\approx12$~mm (i.e., $H/R_p\approx0.75$). The scale bars represent 10~mm.  Both images were edited to enhance the brightness/clarity. The artwork was first presented in \cite{Kiyama2022}.}
    \label{fig:vis}
\end{figure*}

Activating a rubber popper underwater can also create cavitation. Upon activation, an inverted rubber popper quickly returns to its original hemispherical shape. The dynamic is called ``snap-through" instability and is widely studied (e.g., \cite{Pandey2014}) and even used as an actuator for soft robotics \cite{Gorissen2020}. Often times, the popper can jump up to a few meters in the air. Even underwater, the popper dynamics remain very fast. Figure \ref{fig:vis} shows the formation of toroidal cavitation in an aqueous solution upon the slamming of the popper to the substrate. The bubble forms spontaneously within a thin gap between the popper and the substrate and lasts for $\sim O(1)$~ms. 
We note that the entire process seems similar to the cavitation reported upon the underwater collision between a solid object and substrate \cite{DeGraaf2014,DeGraaf2015,Mansoor2016}. In these works, the cavitation onset was explained by either the depressurization upon rebound \cite{Mansoor2014} or the high shear stress \cite{Seddon2012}. Both mechanisms may not be applicable to this particular toroidal cavitation resulting from a snapping popper (figure \ref{fig:vis}).

In this paper, we examine the cavitation phenomena caused by an elastic popper. First, we classify three types of cavitaion and discuss their mechanism of cavitation onset. Systematic experiments suggest that a fast water flow squeezed out from a thin gap between the popper surface and the glass substrate dominates the toroidal cavitation, as implied by a conventional cavitation number \cite{Brennen1995}. We then focus on the morphology of the bubble (i.e., the lifetime and radius) and discuss it through the two-dimensional Rayleigh-Plesset equation. We also adopt the energy balance between the inverted popper and the fully expanded cavitation and discuss the physical meaning of the control parameter to provide a theoretical framework. The present paper provides insights into cavitation mechanics that are a result of fluid-elastic interaction.

\section{Methods}
\subsection{Preliminary Experiments and Observations}

We performed a preliminary experiment to capture overall dynamics (see Appendix for the details). First, a rubber popper was mounted on a 3D-printed platform with an inner diameter of 3~cm. Since the popper was slightly lighter than water and could float, this platform was used to fix the popper's location. The initial height of the platform determines the parameter $H$. The stand-off parameter $H/R_p$ is defined as the initial popper location normalized by the popper radius $R_p$. The stand-off parameter is varied as $0.5\leq H/R_p\leq1.4$. Five experiments were conducted for each condition. 

We used commercially available poppers (ArtCreativity com.) of two different radii, $R_p\sim16$~mm and $\sim$22~mm. We assumed that Young's modulus is $E\sim25$~ MPa based on a previous study \cite{Pandey2014}. We used deionized water as a working fluid, where the density and the vapour pressure were assumed to be $\rho\sim1,000$~kg/m$^3$ and $p_v\sim2$~kPa. In figure \ref{fig:vis}, we used a glycerol-water mixture ($\approx50\%$ by volume) for visualization purposes, whose viscosity is expected to be slightly higher than the water ($\sim5$~cSt \cite{Madison1932}). The experiments were performed in Ithaca, NY. We assumed the atmospheric pressure to be $p_0\sim101$~kPa.

The dynamics of the popper were captured by synchronized high-speed cameras. The bottom-view images were recorded by two Phantom Fastcam NOVA (5,000 frames per second) either directly or through mirrors. The cavitation onset was manually detected and bubble lifetime $t_\mathrm{life}$ and bubble size $R_\mathrm{in}$ and $R_\mathrm{out}$ were estimated. The side-view images were filmed by Photron Fastcam SA-Z (5,000 frames per second). 

\begin{figure*}
    \centering
\includegraphics[width=0.85\paperwidth]{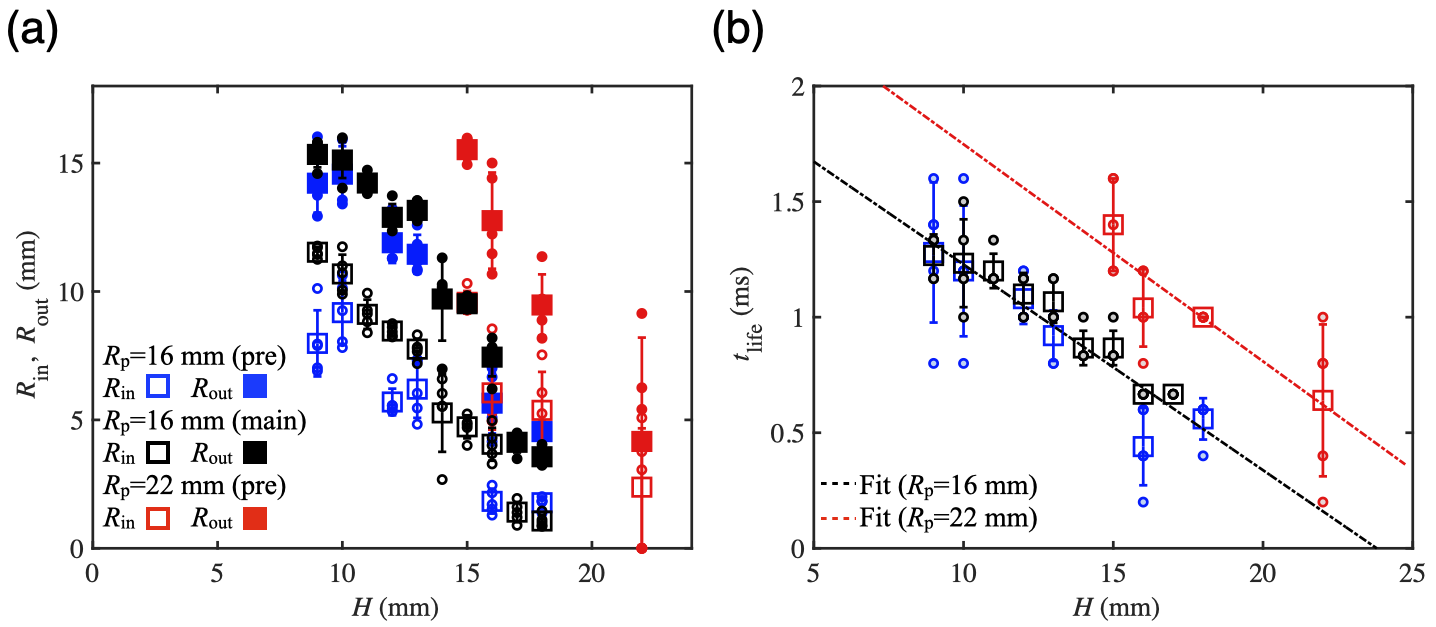}
    \caption{(a) Radius of the cavitation $R_\mathrm{in}$ (open markers) and $R_\mathrm{out}$ (filled markers) as a function of the platform position $H$. Colors distinguish the popper sizes (blue for $R_p=16$~mm in preliminary experiment, black for $R_p=16$~mm in main experiment, and red for $R_p=22$~mm in preliminary experiment). We note that we have five trials at each condition (marked by dots) to obtain the mean value (square) and standard deviation (error bar). (b) Lifetime of the cavitation $t_\mathrm{life}$ as a function of the platform position $H$. Trend lines are $t_\mathrm{life}=-0.089H+2.118$ (R-squared value: 0.7306) for a smaller popper (a black dashed line, preliminary and main experiments combined) and $t_\mathrm{life}=-0.094H+2.687$ (R-squared value: 0.5891) for a larger popper (a red dashed line).}
    \label{fig:raw}
\end{figure*}

We first show the general trend of the size of the cavitation bubble $R_\mathrm{in}$ and $R_\mathrm{out}$ as a function of the platform position $H$ (figure \ref{fig:raw}(a)). Blue and red markers show the data from the preliminary experiment, while the black ones represent the main experiment result as explained in the following section. By definition (see figure \ref{fig:vis}(b)), the outer radius $R_\mathrm{out}$ (filled markers) is always bigger than the inner one $R_\mathrm{in}$ (open markers). But their trend as a function of $H$ remains similar to each other for both popper sizes ($R_p=16$~mm and $R_p=22$~mm). It is noted that a larger popper can create a larger bubble while maintaining a similar downward trend against $H$. The bubble lifetime $t_\mathrm{life}$ also goes with a similar trend against $H$ (figure \ref{fig:raw}(b)). 

The preliminary experiment revealed that the toroidal cavitation we presented in figure \ref{fig:vis} can be observed in only limited experimental conditions. If a popper is released too close to the substrate (i.e., small $H$), the popper could not be accelerated fast enough to cavitate fluid. Indeed, we observed non or only a partial toroidal bubble at $H=7$~mm for $R_p\sim16$~mm (not shown in figure \ref{fig:raw}). Also, the cavitation bubbles except for the ring-type bubble (discussed in the following section) vanish if $H$ is very large. The dimensionless popper location $H/R_p$ seemed to be the primary parameter to describe the phenomena.

\subsection{Main Experiments}
\begin{figure*}
    \centering
\includegraphics[width=0.85\paperwidth]{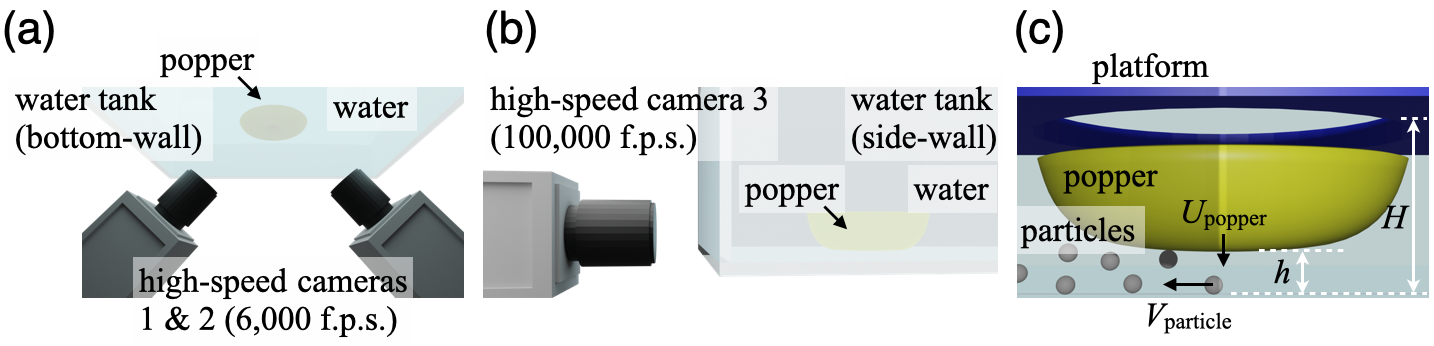}
    \caption{(a) Schematic of the bottom-view measurement by employing two high-speed cameras (not to scale). (b) Schematic of the side-view measurements with and without tracer particles (see also the left- and right-hand side portions of figure \ref{fig:setup}(c), not to scale). Some of the measured quantities ($H, h, U_\mathrm{popper},$ and $V_\mathrm{particle}$) are presented in figure \ref{fig:setup}(c), while they were calculated in separate measurements.}
    \label{fig:setup}
\end{figure*}

After performing the preliminary experiment, it became evident that close-up observations were necessary. We performed the three-dimensional imaging by employing a simpler setup and higher magnification (figure \ref{fig:setup}(a)) to estimate the inner shape and slamming speed of the popper right before the cavitation onset. The frame rate of the two Photron NOVA high-speed cameras was set at 6,000 frames per second. The spatial resolutions were almost identical to each other (25.56 and 27.45 pixels/mm). In this experiment, we selected one popper, whose radius was $R_p\sim16$~mm and whose surface was painted by black dots (see figure \ref{fig:regimes}), as a representation. Image pairs were cross-correlated through a DLTdv8 digitizing tool \cite{Hedrick2008} to estimate the deflection of the inverted popper surface. The software is available for free and can be run as the Matlab Application. Dotted patterns were tracked semi-manually to compute values in not only the $x$ and $y$ but also the $z$ coordinates. The vertical speed of the slamming popper, $U_\mathrm{popper}$, can be estimated as $U_\mathrm{popper}=\Delta z_\mathrm{popper}/\Delta t_1$, where $\Delta z_\mathrm{popper}$ is the vertical displacement of the popper center for a short period of time $\Delta t_1$ before cavitation onset ($\Delta t_1=1$~ms). We tested 10 different $H$ levels (from 9~mm to 18~mm, $0.56\leq~H/R_p\leq1.13$) and repeated measurement 5 times.

We also performed two side-view measurements (see figure \ref{fig:setup}(b)). One of them was to measure the flow speed right before the cavitation onset. Silver-coated ceramic particles with a typical diameter of $d\sim85$~$\mu$m were seeded (see the left-hand side of figure \ref{fig:setup}(c)). A Photron SA-Z high-speed camera could achieve a spatial resolution of $\sim32$~pixels/mm while maintaining a high temporal resolution (100,000 frames per second). We kept using the same popper ($R_p\sim$16~mm) while changing the release height $H$ for 5 different levels (from 9~mm to 18~mm, $0.56\leq~H/R_p\leq1.13$). Particle tracking was performed via the free software Tracker (e.g., \cite{Brown2009}) for 0.3~ms until cavitation starts. As a measure of the flow speed, the radial speed of the particle $V_\mathrm{particle}=\Delta r_\mathrm{particle}/\Delta t_2$ was estimated, where $\Delta r$ is the radial displacement of the particle for a short period of time $\Delta t_2$ before cavitation onset ($\Delta t_2=0.1$~ms). We note that we assume the popper and fluid dynamics were to be axis-symmetric.  While the particle dispersion and popper dynamics are not exactly the same for each trial, the general behavior was confirmed to be similar enough based on 5 trials (see Appendix).

We also filmed the side-view images of the same popper ($R_p\sim$16~mm) slamming the substrate in water without particles, to obtain a better understanding of the thickness of the fluid gap between the popper and substrate, $h$, and that of the bubble (see also the right-hand side of figure \ref{fig:setup}(c)). We used a Photron SA-Z high-speed camera at 100,000 frames per second at 15.6~pixels/mm. The gap $h$ is measured at one frame earlier than the cavitation onset. Experiments were repeated 5 times for each condition (from 9~mm to 18~mm, $0.56\leq~H/R_p\leq1.13$).

In addition, we measured the force $F$ that the popper can induce, to scale the kinetic energy released. We employed a force sensor (DYLY-106 S-type load cell, measurable range: up to 2~kg) and placed it under the popper activating it in the air (Appendix). The force sensor is connected to the amplifier (2310B Signal Conditioner Amplifier, Gain: 2.0$\times10^2$) and the data acquisition system (National Instrument DAQ USB-6001). The data were processed through the Matlab Analog Input Recorder (sampling rate: 20,000~Hz, sampling duration: 10~s) in the PC. We used the same platform to change $H$ while holding the popper by hand until it gets activated. Once the popper is activated, it slams a 3D-printed circular plate that is mounted on the top of the force sensor. The calibration information for the force sensor is shown in the Appendix.

\section{Results and Discussion}
\subsection{Cavitation Mechanism}

\begin{figure*}
    \centering
\includegraphics[width=0.85\paperwidth]{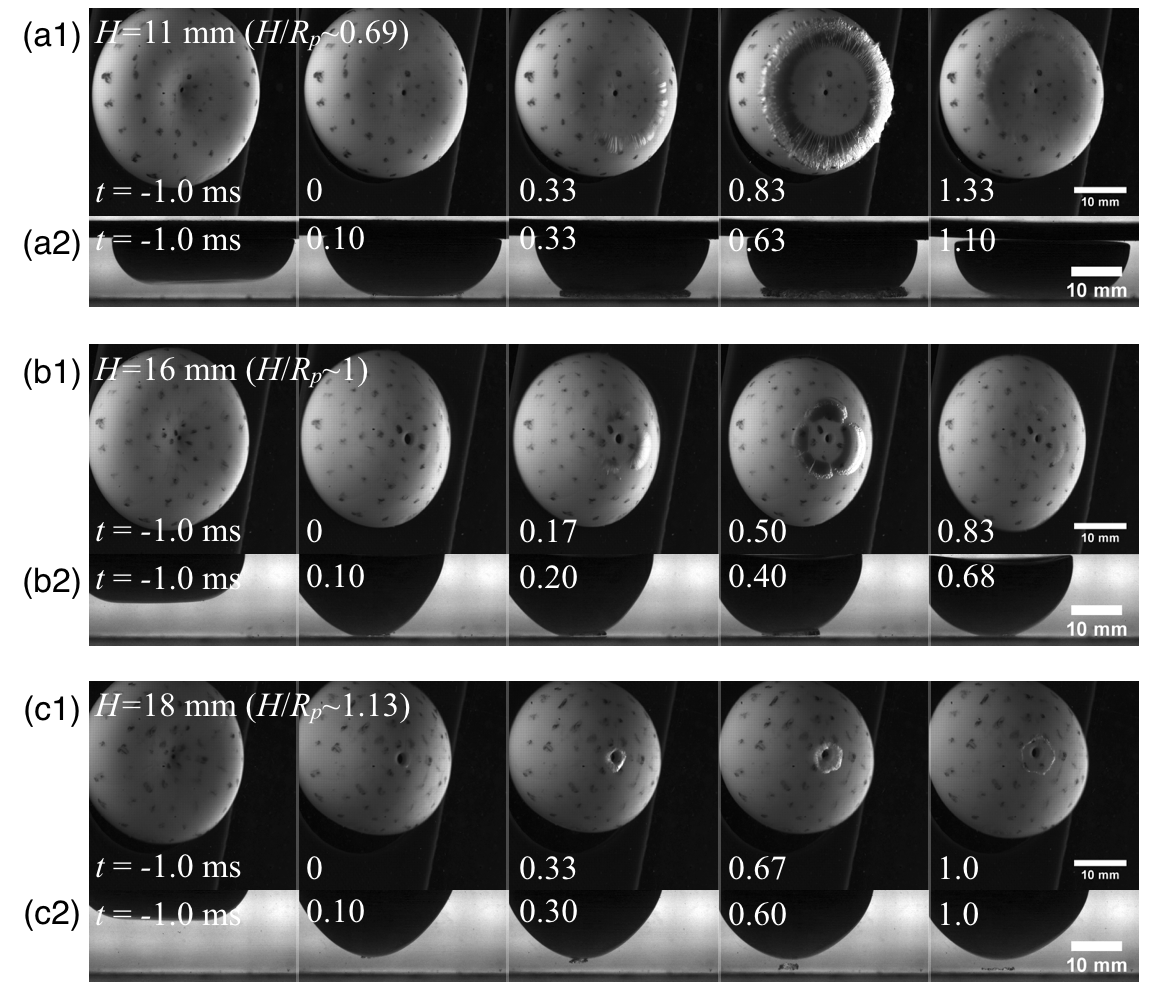}
    \caption{High-speed images of cavitation phenomena upon popper slamming ($R_p\sim$16~mm). (a1) angled bottom-view of the toroidal bubble at $H/R_p\sim0.69$. The dark dots and tiltness are used to measure 3D shapes. The frame rate was 6,000 frames per second. (a2) side-view images of the toroidal bubble separately filmed at the same condition with the frame rate of 100,000 frames per second. (b1 \& b2) high-speed images of the bubble in the transition regime ($H/R_p\sim1$). (c1 \& b2) high-speed images of the vortex-ring type bubble formed at the tip of the popper ($H/R_p\sim1.13$).}
    \label{fig:regimes}
\end{figure*}

We observed three different bubble dynamics: the toroidal cavitation, the vortex-ring type bubble, and the transition. The first to be discussed is the toroidal cavitation that was mentioned in the preliminary observation (figure \ref{fig:vis}). As shown in figure \ref{fig:regimes}(a1), bubbles form from the middle of the popper surface and then expand, maintaining a toroidal shape if $H/R_p\ll$1. The bubble formation seems to be similar to the cavity formation upon a sphere water entry \cite{Marston2012}, where the gas phase expands from the three-phase contact point. From the side, it can be clearly seen that when the bubble begins to form, there is a thin gap between the bottom of the popper and the substrate (see $t=0.1$~ms in figure \ref{fig:regimes}(a2)). We note that the bubble does not form as the symmetric torus when $H/R_p$ is too small, as noted in the preliminary experiment at $H/R_p\sim0.44$. From the bottom view in the main experiment, we observed this partial cavitation when $H/R_p\sim0.56$ as well as in some of the $H/R_p\sim0.63$ trials. The fully-developed toroidal bubble was observed within the range starting from $H/R_p\sim0.63$ up to $H/R_p\sim0.88$. 

We observed another unique bubble at the other end of parameter space (i.e., $H/R_p\gg1$). The bubble formed not from the mid-surface but at the tip of the hole on the popper (see $t=0.33$~ms in figure \ref{fig:regimes}(c1)). The side-view images indicate that the vortex ring-type bubble is ejected from the hole at some transrational speeds and levitates in the gap for a while (figure \ref{fig:regimes}(c2)). The mechanism of bubble onset is different than the toroidal cavitation at a smaller $H/R_p$ and seems to be dominated by the popper dynamics. 

The third regime is the intermediate one. The bubble showed a somewhat ring-like shape (figure \ref{fig:regimes}(b1)) but was not as uniform as the toroidal cavitation. The destruction started to appear from $H/R_p\sim0.88$ as ``cracks" on the bubble surface and becomes apparent when $H/R_p\sim1.0$. The gap $h$ between the popper and the substrate becomes very thin and almost not visible ($t=0.1$~ms in figure \ref{fig:regimes}(b2)). The popper perhaps recovered its original shape as $H/R_p$ approaches 1.0, but still generates cavitation near the center of the popper.

\begin{figure*}
    \centering
\includegraphics[width=0.85\paperwidth]{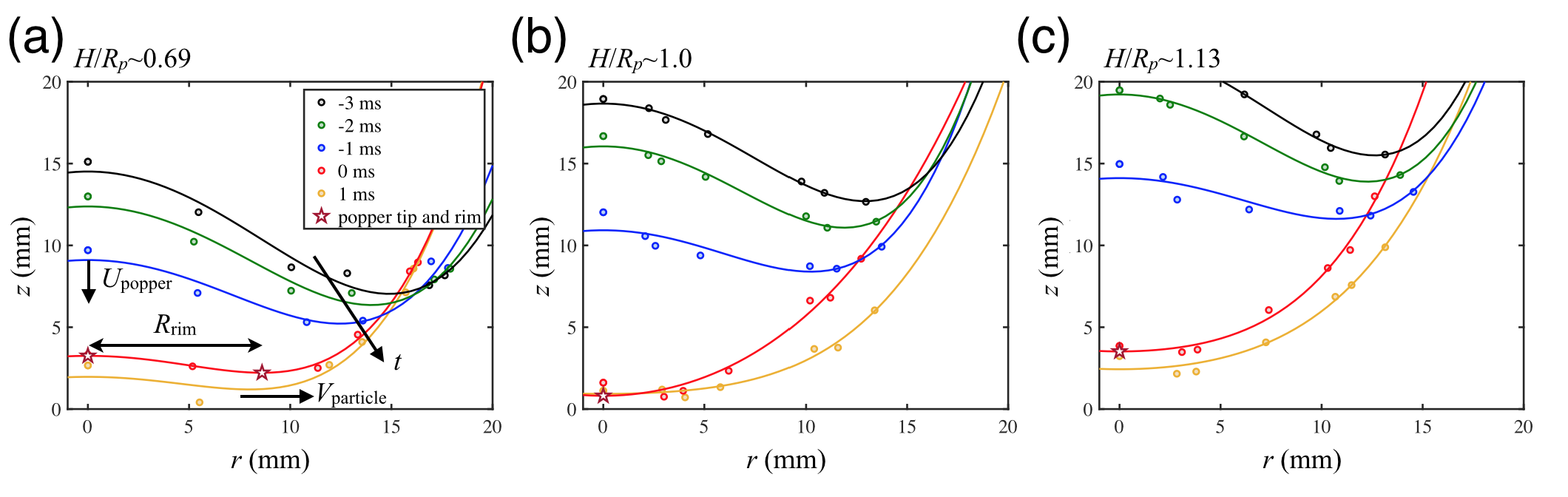}
    \caption{The bottom shape of the popper ($R_p\sim16$~mm) is estimated by the three-dimensional imaging at the frame rate of 6,000 frames per second. Colors distinguish the times from $t=-3$~ms to $t=1$~ms. The reference time $t=0$~ms represents the one frame earlier than the cavitation onset. Solid lines are computed based on the popper height at each distance $r$ (marked by dots) from the center by adopting the fourth-order polynomials while assuming that the popper shape is axis-symmetric. Stars represent the popper center and $R_\mathrm{rim}$ at $t=0$~ms when available. The height of the substrate $z=0$ is set arbitrarily but remains the same for all three panels. The experimental conditions were $H/R_p\sim0.69$ for (a), $H/R_p\sim1.0$ for (b), and $H/R_p\sim1.13$ for (c). We note that the arrows in (a) indicate the direction of the speeds of popper center $U_\mathrm{popper}$ and seeded particles $V_\mathrm{particle}$ that we present in figure \ref{fig:velocity}(a) (see also figure \ref{fig:setup}(c)).}
    \label{fig:poppershape}
\end{figure*}

The three-dimensional imaging enabled us to visualize the complex inner shape of the popper while it is snapping. Figure \ref{fig:poppershape} shows the estimated shape of the popper at different moments (from $t=$-3~ms to $t=$1~ms). The reference time $t=0$~ms represents one frame earlier than the cavitation onset. When $H/R_p\ll1$ (figure \ref{fig:poppershape}(a)), the popper maintains either a flattened or even inverted bottom shape at the time of cavitation. The location of the extreme $R_\mathrm{rim}$ at $t=0$~ms, where the distance between the popper and the substrate becomes minimum, was computed and marked by a star in figure \ref{fig:poppershape}(a). A similar popper bottom shape was observed for $0.56\leq H/R_p\leq0.81$. As $H/R_p$ increases, the popper recovers the hemispherical shape when it approaches the substrate ($t\sim0$~ms). A stretched popper for $H/R_p\sim1.0$ touched (or approached close enough) the substrate as suggested by the slight movement of the popper tip between $t=0$~ms and $t=1$~ms (figure \ref{fig:poppershape}(b)). When $H/R_p\gg1$, the popper is still moving when it induces the vortex-ring type bubble (figure \ref{fig:poppershape}(c), $t=0$~ms and $t=1$~ms). The popper oscillates and travels toward the substrate (see also figure \ref{fig:regimes}(c2)). Comparing the traveling distance of the popper center between $t=-1$~ms and $t=0$~ms in figures \ref{fig:poppershape}(a \& c), it is visible that the speed of the popper increased as $H/R_p$ increased.

Regardless of the bubble dynamics type, the liquid pressure needs to be reduced significantly to cavitate. The cavitation number $Ca$ is a powerful tool to scale the likelihood of cavitation (e.g., \cite{Brennen1995}), which compares the pressure threshold $p_0-p_v$ and pressure drop $\Delta p$ as
\begin{equation}
    Ca=\frac{p_0-p_v}{\Delta p}.
    \label{eq:ca}
\end{equation}
Here, $p_0$ and $p_v$ are respectively the atmospheric ($p_0=101$~kPa) and liquid vapor ($p_v=2$~kPa for water) pressures. This dimensionless number tells us that the lower the Cavitation number, the higher the chance of cavitation. An appropriate representation of $\Delta p$ may vary depending on the mechanism of cavitation \cite{Pan2017}. In this manuscript, we adopted the conventional dynamic pressure representation to estimate it as $\Delta p\sim\frac{1}{2}\rho V^2$, where $V$ is the characteristic flow speed of this expansion flow \cite{White2003}.

\begin{figure*}
    \centering
\includegraphics[width=0.85\paperwidth]{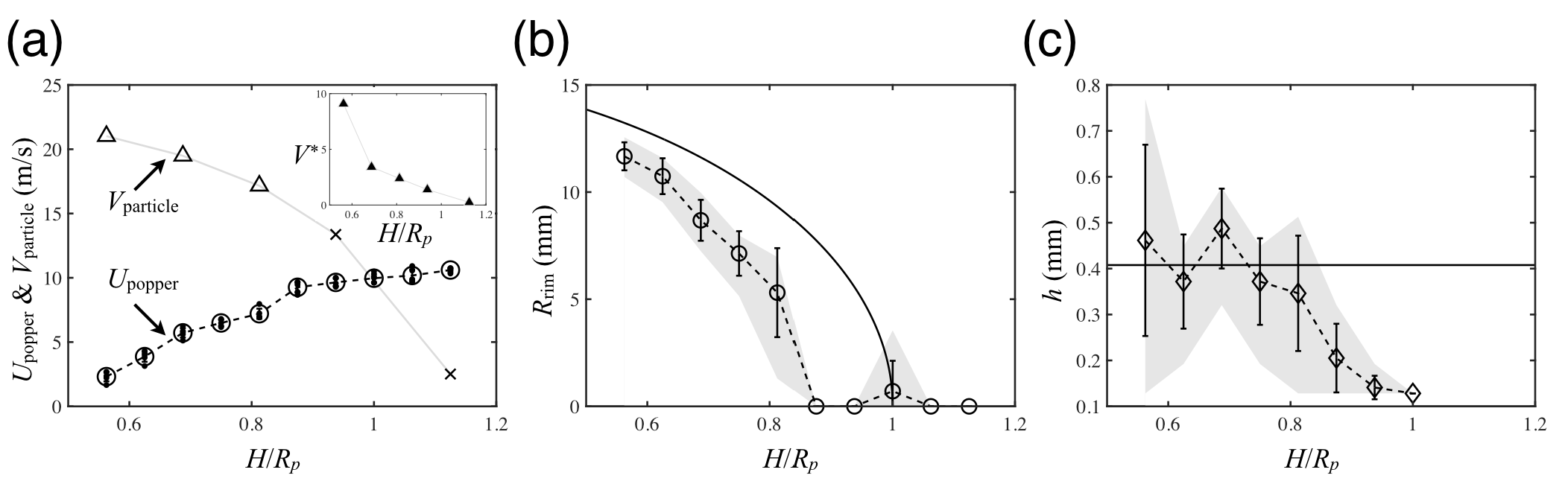}
    \caption{(a) The circles show the vertical velocity of the popper, $U_\mathrm{popper}$, measured through the three-dimensional imaging as shown in figure \ref{fig:poppershape}. The open circles represent the mean value of five trials at the same condition, while small dots show the individual trials. The triangles show the speed of the particles in the radial direction, $V_\mathrm{particle}$, measured through the particle tracking below the popper bottom surface. The fastest particles for a larger $H/R_p$ were found outside of the popper bottom surface ($r>R_\mathrm{rim}$), which are marked by crosses. The inset shows the ratio of speeds $V^\ast=V_\mathrm{particle}/U_\mathrm{popper}$ computed by the values in figure \ref{fig:velocity}(a) as a function of $H/R_p$. (b) The circles show the location of the extreme of the polynomials that fit the popper bottom shape (e.g., figure \ref{fig:poppershape}), $R_\mathrm{rim}$. Markers and error bars respectively show the mean and standard deviation calculated based on five trials. A gray shaded region shows the maximum and minimum values for each condition. A solid line shows the first-order approximation $R_\mathrm{rim}\sim~R_p\sqrt{1-(H/R_p)^2}$. A discrepancy from the data perhaps indicates that the popper stretches more than its original length due to its elasticity. (c) The thickness of the gap $h$ was measured through side-view imaging (marked by the diamonds). Markers and error bars respectively show the mean and standard deviation calculated based on five trials. A gray shaded region shows the maximum and minimum values for each condition. A solid line shows $h\sim0.4$~mm, which was calculated from the data for $0.56\leq H/R_p\leq0.81$. Note that we did not measure $h$ when $H/R_p>1$, as the vortex-ring-type bubbles occurred regardless of the distance from the substrate.}
    \label{fig:velocity}
\end{figure*}

The popper center might move fast enough to cavitate water when $H/R_p$ is large enough. Circles in figure \ref{fig:velocity}(a) show the speed of the popper center, $V_\mathrm{popper}$, which is estimated by the three-dimensional imaging data (see also figure \ref{fig:poppershape}). In general, $V_\mathrm{popper}$ increases as $H/R_p$ increases. It showed a somewhat flat response for a larger $H/R_p$ values, perhaps because the popper achieved the maximum stretch. The popper speed $V_\mathrm{popper}$ reached $V_\mathrm{popper}\sim11$~m/s, which gave us a Cavitation number of $Ca\sim1.64$. Because this $V_\mathrm{popper}$ is the averaged speed over the relatively long time interval ($\Delta t_1\sim1$~ms), we can safely assume that the instantaneous speed is faster, and might satisfy $Ca<1$. It suggests that the vortex-ring type bubble (the third regime, figure \ref{fig:regimes}(c)) occurs due to the fast snapping of the tip of the popper center. 

The toroidal cavitation was observed when $H/R_p\geq0.81$ as discussed earlier., where $V_\mathrm{popper}$, i.e., the measure of the highest speed that the popper can achieve, was not fast enough to cavitate water. In such cases, the flow near the substrate was however fast enough. The result of the particle tracking from the side-view shows that the particles below the flattened popper bottom (marked by the triangles in figure \ref{fig:velocity}(a)) can achieve $Ca<1$ (i.e., $>14.1$~m/s). The speed of the fastest particle ($V_\mathrm{particle}$) slowed down as $H/R_p$ increased. Therefore, we conjecture that the toroidal cavitation occurs because the flow in the gap between the popper and substrate (see also figures \ref{fig:vis}(a) and \ref{fig:regimes}(a2)) is accelerated significantly. Here, let $\Omega=\pi R^2_\mathrm{rim}h$ be a cylindrical fluid volume below the flattened popper bottom. $R_\mathrm{rim}$, which is estimated as the location of the extreme in the fitting curve (see figure \ref{fig:poppershape}(a)), became smaller as $H/R_p$ became larger for $H/R_p\leq0.81$ (marked by circles in figure \ref{fig:velocity}(b)). A volume conservation (i.e., $d\Omega/dt=2\pi R_\mathrm{rim}h(dR_\mathrm{rim}/dt)+\pi R^2_\mathrm{rim}(dh/dt)=0$) gives us a scaling $V_r\sim(R_\mathrm{rim}/(2h))U_\mathrm{popper}$. This simple scaling law indeed captures the extremely fast flow speed expected from that of particles (figure \ref{fig:velocity}(a)). While the fine scale of the gap $h$ makes it challenging to discuss its trend, our data shows the gap could be $h\sim0.4$~mm for $H/R_p\leq0.81$ (a gray line in figure \ref{fig:velocity}(c)). It implies a flow speed can be as fast as $V_r\sim(R_\mathrm{rim}/(2h))U_\mathrm{popper}\sim30-50$~m/s, which is supposed to be fast enough to cavitate water. We note that the particle speed $V_\mathrm{particle}$ is the measure of the lower bound of the flow speed as discussed in Appendix. The inset of figure \ref{fig:velocity}(a) shows the ratio of speeds $V^\ast=V_\mathrm{particle}/U_\mathrm{popper}$ as a function of $H/R_p$. It shows the radial flow is enhanced with respect to the vertical popper motion at a small $H/R_p$ while qualitatively obeying the scaling. Observations above agree with our hypothesis that the fast flow squeezed out from the thin gap between the popper and substrate drives the toroidal cavitation.

We note that the cavitation in the transition region might occur slightly differently than the toroidal cavitation. $R_\mathrm{rim}$ could not be computed and thus the popper perhaps reached the substrate. The scaling for the water flow in the gap no longer holds. It was in the same line with our side-view visualization that showed the gap $h$ for $H/R_p\geq1$ was not visible or was negligibly small (figure \ref{fig:velocity}(c)). In the particle tracking, the fastest particles were found outside of the popper bottom surface (i.e., $r>R_\mathrm{rim}$, marked by the crosses in figure \ref{fig:velocity}(a)). It suggests that the radial removal of surrounding fluid play a role in cavitation onset in the transition regime, where the detailed mechanism is yet unclear.

It is also important to compare this unique phenomenon to those reported in similar settings. In the previous study, the cavitation bubbles formed immediately after the collision of the rigid sphere were spherically nucleated around the impact point \cite{Mansoor2014}.
In contrast, the bubbles in our study nucleate annually without physical contact between the popper and the substrate. We do not observe bubbles in the central region, indicating that pressure reduction is localized in the annulus region, which is perhaps assisted by the snap-through dynamics of the popper. The bubbles in the annulus then merge to form a toroidal bubble during the evolving stage (figure \ref{fig:vis}(b), $t=0.33~-~0.89$~ms). We also note that our experiment does not provide sufficient evidence to determine the contribution of stress-induced cavitation \cite{Seddon2012}. The shear stress might be scale as $\sigma\sim\mu(\partial V_r/\partial z)\sim\mu(V_r/h)$ by assuming a Couette flow. This simple scaling predicts $\mu(V_r/h)\sim50$~Pa~$\ll(p_0-p_v)$ for the toroidal cavitation cases, where $\mu\sim1$~mPa$\cdot$s, $V_r\sim20$~m/s and $h\sim$0.4~mm are assumed. However, it might become dominant in the transition regime (the second regime, $H/R_p\sim1$) as the gap $h$ can be extremely thin (figure \ref{fig:velocity}(c)).

\subsection{Cavitation Morphology}
We first discuss how the lifetime of the cavitation bubble is related to its size through the equation of motion (i.e, Rayleigh-Plesset equation \cite{Rayleigh1917,Plesset1949}). For simplicity, we make a crude assumption that bubble dynamics are two-dimensional and purely radial. We also assume that the inner radius ($R_\mathrm{in}$) does not move within the lifetime of the bubble. These assumptions imply that we derive a scaling law for cavitation bubbles in the toroidal cavitation and the transition regimes, where the bubble dynamics are largely restricted to the two-dimensional. We note that the vortex-ring-type bubble requires the translational speed taken into account \cite{Chahine1983}, which is not the scope of this study.
We may use the two-dimensional Rayleigh equation (e.g., \cite{Lohse2004}) in terms of the bubble radius $R$, as
\begin{equation}
    \bigg[\bigg(\frac{dR}{dt}\bigg)^2+R\frac{d^2R}{dt^2}\bigg]\log\bigg(\frac{R}{R_{\infty}}\bigg)+\frac{1}{2}\bigg(\frac{dR}{dt}\bigg)^2=\frac{p_v-p_0}{\rho}.
\end{equation}
We neglected the influence of viscosity, surface tension, and dissolved gas. With the approximation of $\log(R/R_{\infty})\approx1$ and $R~(d^2R/dt^2)+3/2(dR/dt)^2\approx~(d^2R^2/dt^2)/2$ \cite{Duclaux2007}, the equation above can be rewritten as
\begin{equation}
    \frac{d^2R^2}{dt^2}\approx~2~\bigg(\frac{p_v-p_0}{\rho}\bigg).
\end{equation}
We solve this equation in terms of the bubble collapse stage to estimate the characteristic timescale $\tau$.
We use the initial conditions $R=R_\mathrm{out}$ and $dR/dt=0$ at $t=0$, and then obtain
\begin{equation}
    R^2=R^2_\mathrm{out}+\bigg(\frac{p_v-p_0}{\rho}\bigg)~t^2.
\end{equation}
The timescale $\tau$, that a bubble requires to shrink from $R=R_\mathrm{out}$ at $t=0$ to $R=R_\mathrm{in}$ at $t=\tau$, can be scaled as 

\begin{equation}
    \tau~\sim\sqrt{(R^2_\mathrm{out}-R^2_\mathrm{in})~\frac{\rho}{p_0-p_v}}.
    \label{eq:tlife}
\end{equation}

Note that this becomes compatible with the three-dimensional Rayleigh-type bubble lifetime if $R_\mathrm{in}=0$.


\begin{figure*}
    \centering
\includegraphics[width=0.85\paperwidth]{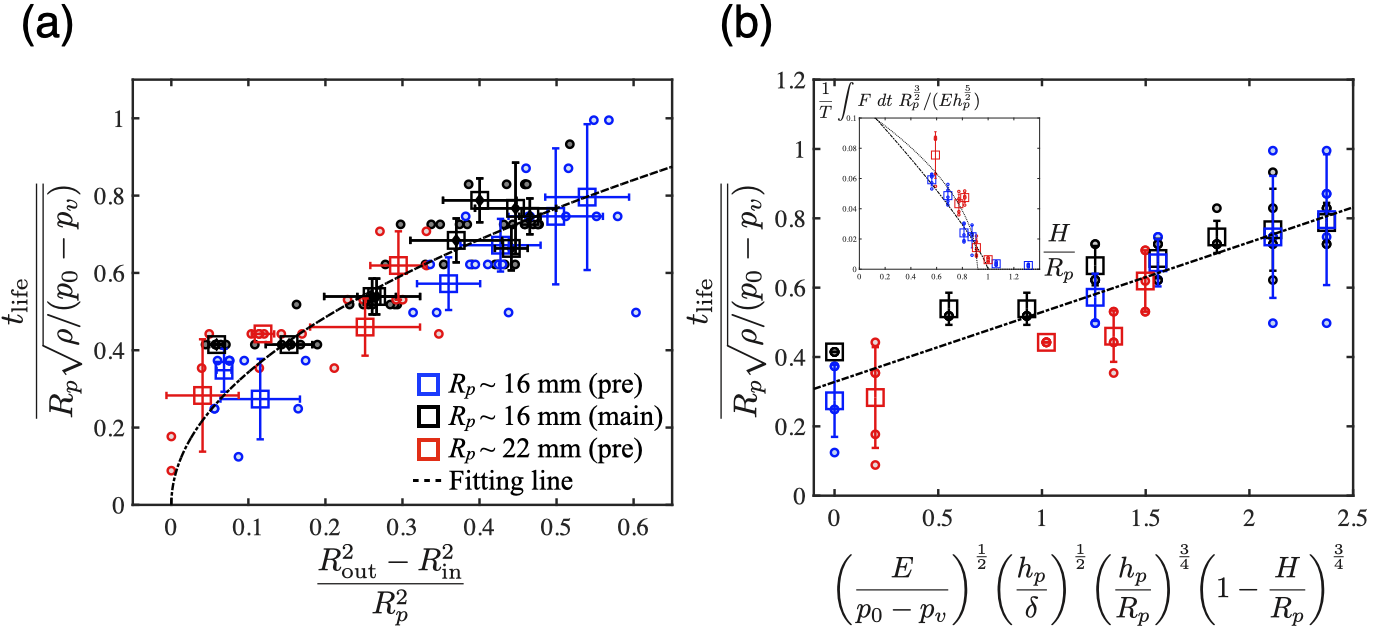}
    \caption{Comparison between the normalized bubble lifetime $t_\mathrm{life}/(R_p\sqrt{\rho/(p_0-p_v)})$ versus the bubble size $(R^2_\mathrm{out}-R^2_\mathrm{in})/R^2_p$. The same data set was used as that in figures \ref{fig:raw}. Squares represent the mean value over five trials that are marked by dots. The error bars are showing the standard deviation. Colors distinguish the typical popper size at rest. The fitting line shows a trend, where the equation was estimated to be $t_\mathrm{life}/(R_p\sqrt{\rho/(p_0-p_v)})=1.085~[(R^2_\mathrm{out}-R^2_\mathrm{in})/R^2_p]^{\frac{1}{2}}$ (R-squared value: 0.6994). (b) Comparison between the normalized bubble lifetime and the timescale based on the elastic potential energy, which was found in equation \ref{eq:final}. The same representation of the data as that in (a) holds. The fitting line is found to be $t_\mathrm{life}/(R_p\sqrt{\rho/(p_0-p_v)}=0.202[E/(p_0-p_v)]^\frac{1}{2}(h_p/\delta)^\frac{1}{2}(h_p/R_p)^\frac{3}{4}(1-H/R_p)^\frac{3}{4}+0.328$ (R-squared value: 0.670). An inset shows the comparison between the normalized force $\frac{1}{T}\int F~dt~R_p^\frac{3}{2}/(Eh^\frac{5}{2}_p)$ (measured in the air) versus the stand-off parameter $H/R_p$. The experimental procedure is shown in the Appendix. Squares represent the mean value over five trials that are marked by dots. The error bars are showing the standard deviation. Colors distinguish the typical popper size at rest. The dotted line shows a fitting $F^\prime\sim 0.111~(0.92-H/R_p)^\frac{1}{2}$ (R-squared value: 0.736). The dashed line shows $F^\prime\sim 0.136(1-H/R_p)^{0.83}$ (R-squared value: 0.786).}
    \label{fig:tlife3}
\end{figure*}

Equation \ref{eq:tlife} indeed describes the experimental data well (figure \ref{fig:tlife3}(a)), despite the simplifications we made. Both quantities are measured experimentally and normalized by the Rayleigh-type factor $R_p\sqrt{\rho/(p_0-p_v)}$. Colors represent the difference in the popper size $R_p$. Red and blue markers show data from the preliminary experiment, while black markers represent the ones from the main experiment. Squares represent the mean values over the five trials, while the error bars show the standard deviation. Individual trials were marked by dots. It is visible that data series collapsed well and showed an incremental trend. The best fit for these data (solid line) $t_\mathrm{life}/(R_p\sqrt{\rho/(p_0-p_v)})=1.085~[(R^2_\mathrm{out}-R^2_\mathrm{in})/R^2_p]^\frac{1}{2}$ scales the general behaviour for both popper sizes, while the prefactor differed from 2. This supports our approach that employing the simplified Rayleigh-Plesset-type model to capture the morphology of the toroidal cavitation bubble. We note that a few data points are showing non-zero $t_\mathrm{life}$ values while $(R^2_\mathrm{out}-R^2_\mathrm{in})/R^2_p=0$ in figure \ref{fig:tlife3}. In these cases, the bubbles formed partially but not fully developed annual shapes at small $H/R_p$ and thus the radii were not identified. We note that the bubble at $H/R_p\sim1.13$ was clearly the vortex-ring type one (figure \ref{fig:regimes}(c)) and thus excluded from the plot. 


To connect the bubble and the popper dynamics, we consider the energy balance between them. For simplicity, we assume that an elastic potential energy, $Y_\mathrm{elastic}$, which is stored in the indented hemispherical shell, will be fully used to form a toroidal bubble and thus be balanced with the hydrostatic potential energy of the bubble at its maximum size. A simple energy balance can yield
\begin{equation}
    (R^2_\mathrm{out}-R^2_\mathrm{in})\sim\bigg(\frac{Y_\mathrm{elastic}}{\pi~\delta(p_0-p_v)}\bigg),
    \label{eq:size}
\end{equation}
where $\delta$ is the characteristic thickness of the toroidal bubble. We note that we were not able to measure $\delta$ precisely. Here, we arbitrarily set $\delta=1$~mm, which is slightly larger than the gap thickness $h$ but still smaller than the outer rim of the fully expanded toroidal bubble. The elastic potential energy $Y_\mathrm{elastic}$ can be approximated through the indentation force $F$ \cite{Audoly2010} as
\begin{equation}
    Y_\mathrm{elastic}\sim \int F~dx\sim\frac{2}{3}\frac{E~h^\frac{5}{2}_p}{R_p}e^\frac{3}{2}\sim\frac{2}{3}\frac{E~h^\frac{5}{2}_p}{R_p}(R_p-H)^\frac{3}{2}.
    \label{eq:uelastic}
\end{equation}
Parameters $E$, $h_p$, and $e$ are Young's modulus, the characteristic thickness of the popper, and the depth of indentation, respectively. The indentation depth $e$ is scaled as $e\sim (R_p-H)$ based on the first-order geometrical consideration (neglecting the stretch of the popper) with the initial height of the platform, $H$. We note that we assumed the Young's modulus $E$ and the popper thickness $h_p$ to be constants for simplicity. Hereafter, we use the Young's modulus of $E=25~$MPa from the literature \cite{Pandey2014} and the popper thickness of $h_p\approx$3~mm measured at the rim of the cut popper, although $h_p$ varies slightly along its arc. The uncertainty associated with the choice of $\delta, E$, and $h_p$ would result in the limitation of this approach and thus their influence deserves further investigation. In equation \ref{eq:uelastic}, it is visible that the parameter $H/R_p$ governs the elastic potential energy $Y_\mathrm{elastic}$. Plugging equations \ref{eq:size} and \ref{eq:uelastic} into equation \ref{eq:tlife} would finally gives us the relationship as
\begin{equation}
    t_\mathrm{life}~\sim~\sqrt{\frac{Y_\mathrm{elastic}}{\pi\delta(p_0-p_v)}\frac{\rho}{p_0-p_v}}~\rightarrow~\frac{t_\mathrm{life}}{R_p\sqrt{\rho/(p_0-p_v)}}~\sim~\sqrt{\frac{E}{p_0-p_v}\frac{h_p}{\delta}\bigg(\frac{h_p}{R_p}\bigg)^\frac{3}{2}\bigg(1-\frac{H}{R_p}\bigg)^\frac{3}{2}}.
    \label{eq:final}
\end{equation}

Equation \ref{eq:final} implies that the lifetime of the bubble $t_\mathrm{life}$ is largely determined by the popper geometry $(h_p/R_p)$, which makes sense as the popper size is a parameter that determines both the bubble size and the lifetime (figure \ref{fig:raw}). It is also implied that the location of the popper $(H/R_p)$ is another important parameter. This is intuitive as the larger $H/R_p$ allows the popper to release more energy, which is evidenced by the incremental trend of $U_\mathrm{popper}$ over $H/R_p$ (figure \ref{fig:velocity}). Figure \ref{fig:tlife3}(b) evaluates the equation \ref{eq:final}. All the parameters $E$, $h_p$, $\delta$, and $R_p$ are chosen as mentioned. The dashed line is the best-fit line $t_\mathrm{life}/(R_p\sqrt{\rho/(p_0-p_v)}=0.202[E/(p_0-p_v)]^\frac{1}{2}(h_p/\delta)^\frac{1}{2}(h_p/R_p)^\frac{3}{4}(1-H/R_p)^\frac{3}{4}+0.328$. 
Despite the uncertainties mentioned above, the model captures the incremental trend of the bubble lifetime. This suggests that the bubble lifetime is scaled as a function of the popper dynamics. In other words, the parameter $H/R_p$, which is a measure of the the intensity of the interaction between the popper and the substrate, can dominate the bubble morphology.


The inset in figure \ref{fig:tlife3}(b) compares the normalized force, $F^\prime\sim\frac{1}{T}\int F~dt~R_p^\frac{3}{2}/(Eh^\frac{5}{2}_p)$ measured in the air, as a function of the stand-off parameter $H/R_p$. We note that we employ the impulse ($\frac{1}{T}\int F~dt$) as a measure of the force to capture the overall behaviour. It is visible that the force decreases as $H/R_p$ increases, and reaches a near-zero value at $H/R_p\sim1$, which makes sense based on the geometrical constraint. A dotted line shows a trend line $F^\prime\sim 0.111(0.92-H/R_p)^\frac{1}{2}$. Though a threshold 0.92 was an arbitrary choice to fit the data with the slope of 1/2, it is possibly justifiable because both the platform and the popper can deform and the threshold can differ from $H/R_p=1.0$. For comparison purposes, the dashed line denotes the best fit when we restrict the threshold to be $1.0$, where we found $F^\prime\sim 0.136(1-H/R)^{0.83}$. In general, the downward trend of the data implies that $H/R_p$ can control the intensity of the interaction between the popper and the substrate as argued above.


\section{Conclusion}
We demonstrated that the underwater slamming of a rubber popper toy against a glass substrate can induce a toroidal cavitation bubble (figures \ref{fig:vis} and \ref{fig:regimes}). A series of experiments (figure \ref{fig:setup}) indicated that the toroidal cavitation occurs due to a fast liquid flow squeezed out from a thin gap between the rubber popper and the glass substrate (figures \ref{fig:regimes}--\ref{fig:velocity}). As the initial position of the rubber popper in the experiment ($H/R_p$) increased, the bubble dynamics transient from the toroidal one to the vortex-ring type one (figures \ref{fig:regimes}(b \& c)). The bubble lifetime ($t_\mathrm{life}$) and radii ($R_\mathrm{in}$ and $R_\mathrm{out}$) we found for the toroidal cavitation and the transient one to be interrelated through the two-dimensional Rayleigh-Plesset-type model (figure \ref{fig:tlife3}(a)). We also discussed an analytical framework for this uniquely formed cavitation through the energy balance between the deformed rubber popper and the fully expanded cavitation bubble. The parameter $H/R_p$, which scales the elastic potential energy used to form cavitation, captured the qualitative trend of both the popper and the bubble dynamics (figure \ref{fig:tlife3}(b)). This paper might provide a platform for further studies on bubbles formed in a complex system with the involvement of elastic structures that may include the Mantis Shrimp fist \cite{Patek2005} or the brain system \cite{Lang2021}.

\section{Acknowledgments}
This work was supported by NSF grant CBET-2002714 and CMMI-1238169.

\section{Author contributions}
S.J. conceptualized the work; A.K. designed the experiments; A.K. and S.W. conducted the experiments and analyzed data. A.K. wrote the original draft of the manuscript, and S.W. and S.J. edited the manuscript.

\section{{Data avaiability}} 
Most figure files and matlab figure data are available on the Open Science Framework (DOI 10.17605/OSF.IO/YCNV8).

\appendix

\section{Details on the experiment}
\subsection{Preliminary experiment}
\begin{figure*}
    \centering
\includegraphics[width=0.75\paperwidth]{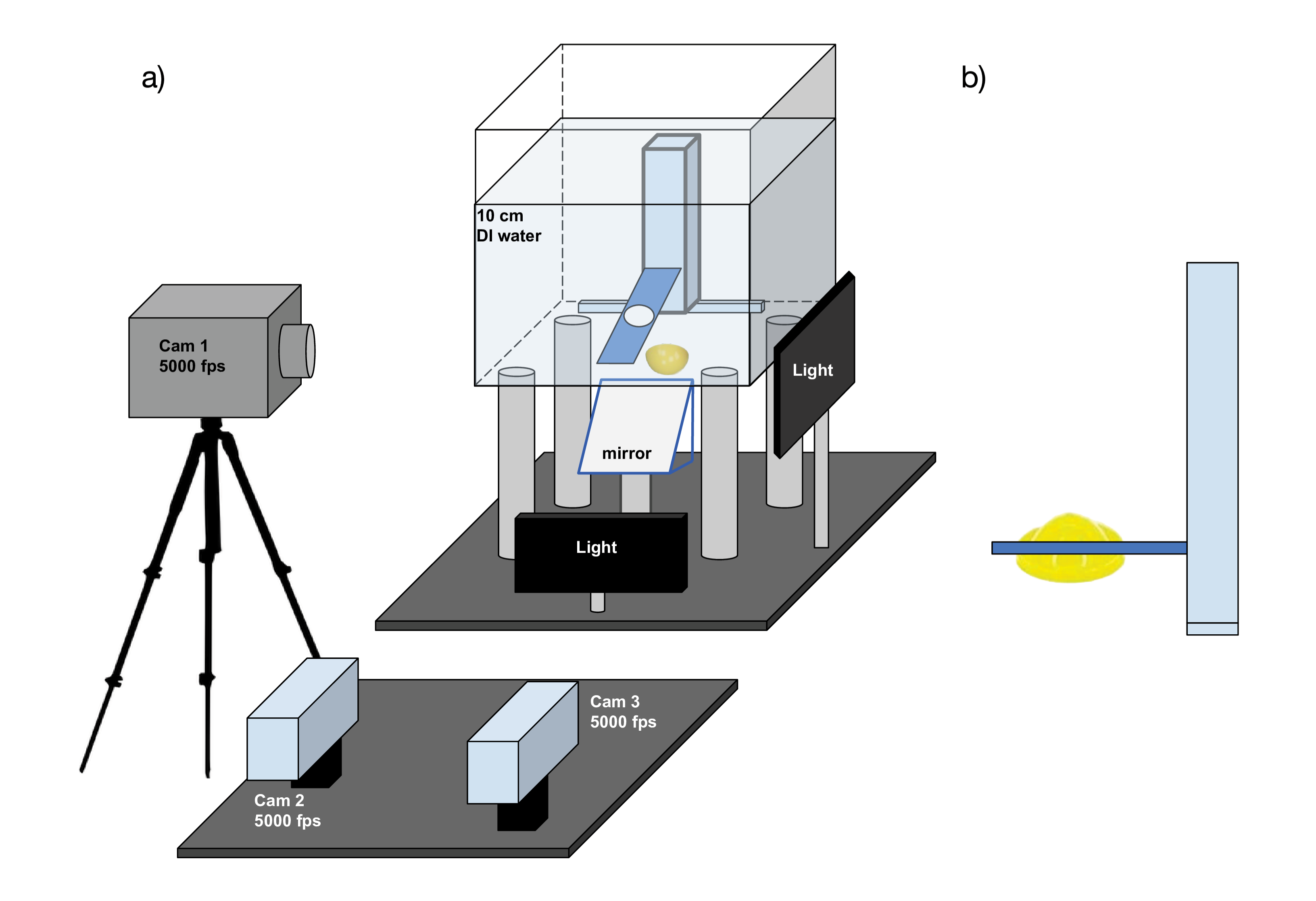}
     \caption{(a) Glass tank is elevated and filled with $\approx 10$~cm of deionized water. A cantilever structure including a 3D-printed platform is submerged in water to position the rubber popper. Three synchronized high-speed cameras are positioned to capture the side-view and bottom-view videos. The frame rate was set at 5,000~frames per second. for all three cameras. (b) 3-D printed platform used to fix popper height. The initial height of the platforms determines the parameter \textit{H}.} \label{fig:pre-setup}
     \end{figure*}

The preliminary experiment was performed by employing three high-speed cameras synchronized at the frame rate of 5,000~frames per second (figure \ref{fig:pre-setup}(a)). The depth of deionized water was set at approximately 10~cm. A 3D-printed platform (see figure \ref{fig:pre-setup}(b)) was submerged and held by hand.

The onset and the collapse of cavitation are visually detected. Thus, the presence of cavitation in this article is defined as the presence of any bubbles, which are larger than a few pixels on the image. The measured data may contain $\pm$1~frame uncertainty that depends on the individual. The size and the lifetime of the bubble are estimated by using the ``reslice” function implemented in the freely available software ImageJ. We assumed the angle made by the two bottom cameras is small enough. We thus directly analyzed either one of these images. The experimental conditions covered were summarized in table \ref{parameters}. 
We note that we only focused on the first onset of the bubble, while secondary cavitation was found to occur.

     
\subsection{Main experiments}

The three dimensional imaging data were correlated through multiple checkerboard images. The vertical ($z$) axis was determined based on the path of the popper center, which was estimated from one of the experimental data for $H/R_p=1.13$. The three dimensional imaging data were also used to estimate the bubble radii $R_\mathrm{in}$ and $R_\mathrm{out}$. Due to the restricted angle between two high-speed cameras, we manually measured the typical bubble radii for each trial through a DLTdv8 digitizing tool \cite{Hedrick2008}. We calculated radius from the popper center as $r=\sqrt{x^2_i+y^2_i}-\sqrt{x^2_0+y^2_0}$ at each time step $i$ (subscript 0 represents the popper center), while the $z$-direction values remained almost constant.

\begin{figure*}
    \centering
\includegraphics[width=0.85\paperwidth]{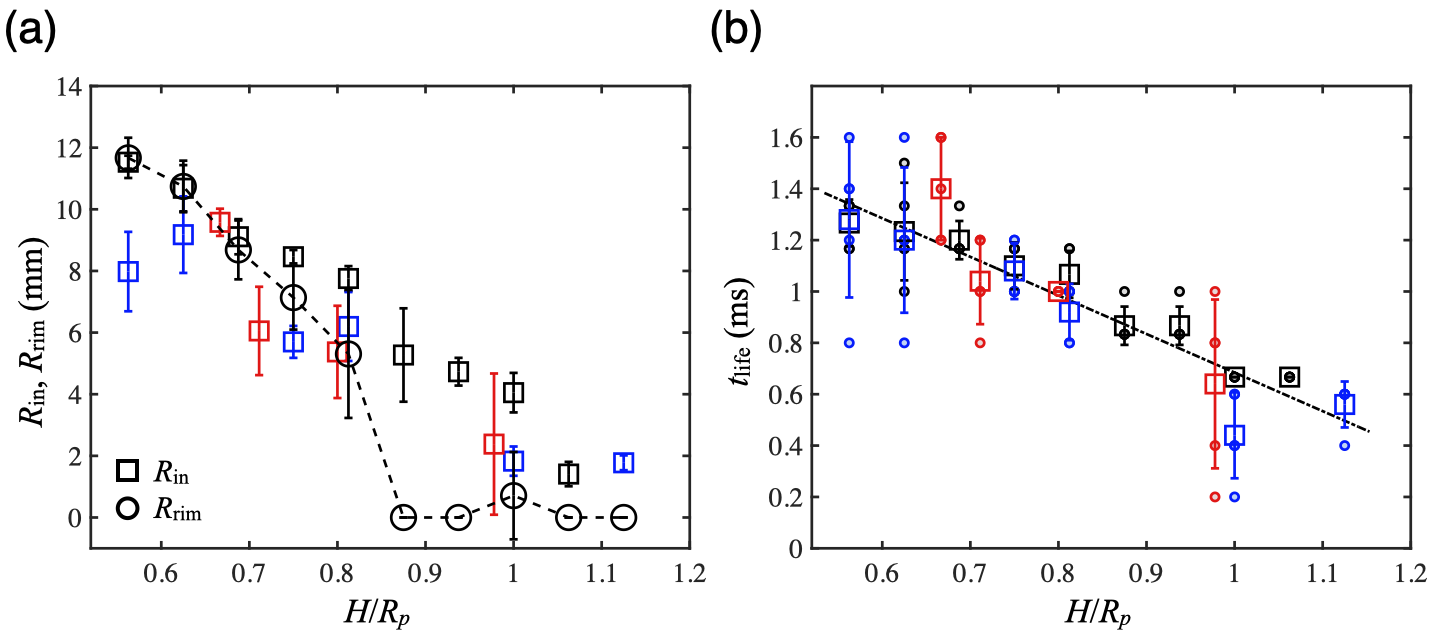}
    \caption{(a) A comparison between the inner radius $R_\mathrm{in}$ (squares, see figure \ref{fig:raw}(a)) and the location of the extreme of the polynominals for the popper bottom shape $R_\mathrm{rim}$ (circles, see figure \ref{fig:poppershape}(b)) as a function of the platform height $H/R_p$. Both radii match well as long as cavitation maintains the toroidal shape ($H/R_p\leq0.81$). (b) The lifetime of the bubble as a function of $H/R_p$ with a trend line of $t_\mathrm{life}=-1.502~(H/R_p)+2.186$ (R-squared value: 0.683).}
    \label{fig:suppl}
\end{figure*}

The inner radius $R_\mathrm{in}$ (squares) is compared with the lowest $r$-direction value of the estimated popper shape $R_\mathrm{rim}$ (circles) (figure \ref{fig:suppl}(a)). The black squares and circles are estimated from the same data set, while red and blue squares are shown for comparison purposes. When the bubble maintains the toroidal shape ($H/R_p\leq0.81$), the inner radius $R_\mathrm{in}$ is pretty close to, or slightly larger than, $R_\mathrm{rim}$, suggesting that a fast radial flow developed within a thin gap nucleated cavitation and expel it outside. In a transition region, a somewhat radial bubble was formed while $R_\mathrm{rim}$ was not well identified ($H/R_p\leq0.88$). We also show the bubble lifetime $t_\mathrm{life}$ as a function of $H/R_p$ for both preliminary and main experiments (figure \ref{fig:suppl}(b)). It shows the bubble dynamics in general were well reproduced for various trials. We note that the trend line here was for comparison purposes and was not based on the theoretical background discussed in the paper.

\begin{table}
\begin{center}
{  \begin{tabular}{|p{1.6cm}|p{2.0cm}|p{2.0cm}|p{4.0cm}|p{4.0cm}|}
\hline
$R_p$ (mm) & $H$ (mm) & $H/R_p$ & Number of conditions $\times$ trials & Experiment type\\[3pt]
\hline
16 & 7 -- 18 & 0.438 -- 1.13 & 7$\times$5 &\multirow{2}{*}{Preliminary}\\
22 & 12 -- 22 & 0.545 -- 1.00 & 5$\times$5 &\\
\hline
16 & 9 -- 18 & 0.563 -- 1.13 & 10$\times$5 & Main (bottom view)\\
16 & 9 -- 18 & 0.563 -- 1.13 & 5$\times$5 & Main (particle tracking)\\
16 & 9 -- 18 & 0.563 -- 1.13 & 10$\times$5 & Main (side view)\\
\hline
16 & 9 -- 21 & 0.563 -- 1.31 & 6$\times$5 &\multirow{2}{*}{Force measurement in air}\\
22 & 13 -- 22 & 0.591 -- 1.00 & 5$\times$5 &\\
\hline
    \end{tabular}
    }
\\
\caption{Summary of experimental conditions}
\label{parameters}
\end{center}
\end{table}

\begin{figure*}
    \centering
\includegraphics[width=0.85\paperwidth]{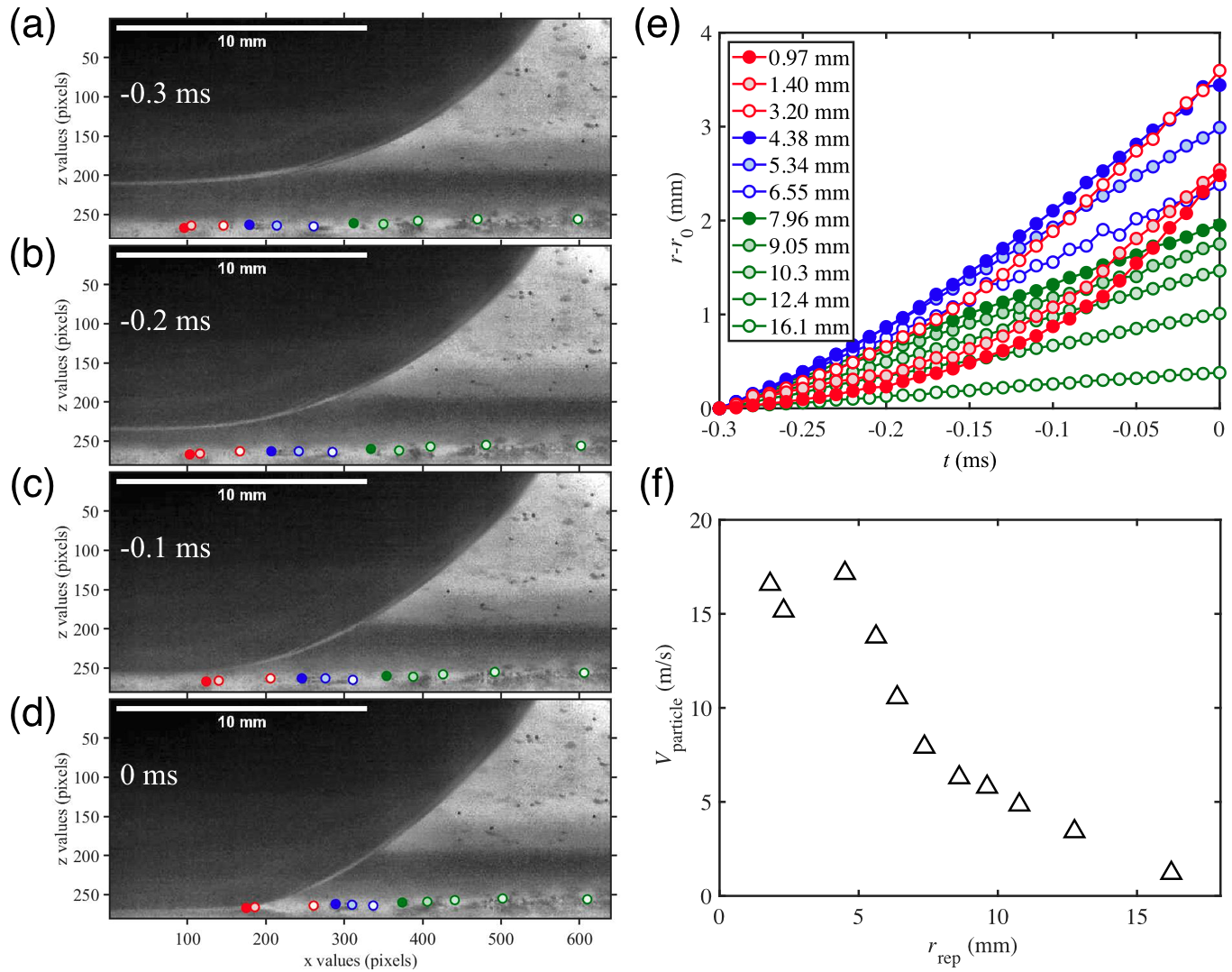}
    \caption{(a-d) Zoom-in images of a popper ($R_p\sim16~$mm) released from a height of $H\sim13$~mm. Relative time with respect to the cavitation onset was -0.3~ms (a), -0.2~ms (b), -0.1~ms (c), and 0~ms (d). The silver-coated ceramic particles ($d\sim$85~$\mu$m) mixed in the deionized water enabled particle tracking. The analyzed particles are marked by dots, where the colors are consistent throughout the images. We note that both axes are in pixels to imply the resolution of the image ($\sim$32~pixels/mm). (e) the displacement of the particles from their original position at $t=-0.3$~ms. Numbers in the legend represent the characteristic particle position $r_\mathrm{rep}$ over the dynamic, which is measured at $t=-0.05$~ms. Colors convey the same meanings as those in (a-d). The averaged particle velocities over 0.1~ms before the cavitation onset ($V_\mathrm{particle}$) as a function of the particle position $r_\mathrm{rep}$.}
    \label{fig:PTV}
\end{figure*}

We adopted the particle tracking approach to estimate the speed of the flow within a narrow gap between the popper and the substrate, which was expected to achieve $V_\mathrm{particle}\sim O(10)$~m/s to satisfy $Ca<1$. We arbitrarily selected the traceable particles out of 5 trials for 5 different $H$ values. 
Figure \ref{fig:PTV} shows the typical images of the particle tracking, where the particles tracked are marked by dots (the height $H$ was set to $H=13$~mm). In this example, we tracked particles at 11 different locations as marked. We note that we selected the particles floating on the glass substrate to estimate the radial flow speed. We acknowledge that the particle motion might be affected by the viscous boundary layer over the substrate and be decelerated. This analysis reflects the lower bound of the flow speeds. As time progresses, particles move to the right, suggesting that the flow is indeed squeezed away from the popper center. This flow pattern is always the case for multiple trials with different $H$ values.

Figure \ref{fig:PTV}(e) shows the time series of the particle positions. The vertical axis represents the displacement of the particles with respect to their original location at $t=-0.3$~ms. The numbers in the legend show the characteristic location of each particle $r_\mathrm{rep}$, which is measured at $t=-0.05$~ms to reflect both their location and speed. Particles located far away from the popper center (green markers) travel at an almost constant speed. The traveling speed of particles decreases as $r_\mathrm{rep}$ increases (see blue and green markers). Interestingly, the particle trajectory for particles closer to the popper center shows a slightly different trend (red markers). Particles do not move much at first (-0.3~ms$\leq~t\leq-0.1$~ms) and then accelerated rapidly ($t\geq-0.1$~ms). We set $\Delta t_2$=0.1~ms to calculate $V_\mathrm{particle}$. Figure \ref{fig:PTV}(f) shows the averaged particle speeds for 0.1~ms $V_\mathrm{particle}$ as a function of $r_\mathrm{rep}$. The speed of outside particles decreases as the distance increases, while inner particles maintain faster speeds. Figure \ref{fig:PTV}(e) \& (f) suggest that the flow reaches at least $V_r\sim$18~m/s, causing cavitation. 

\begin{figure*}
    \centering
\includegraphics[width=0.85\paperwidth]{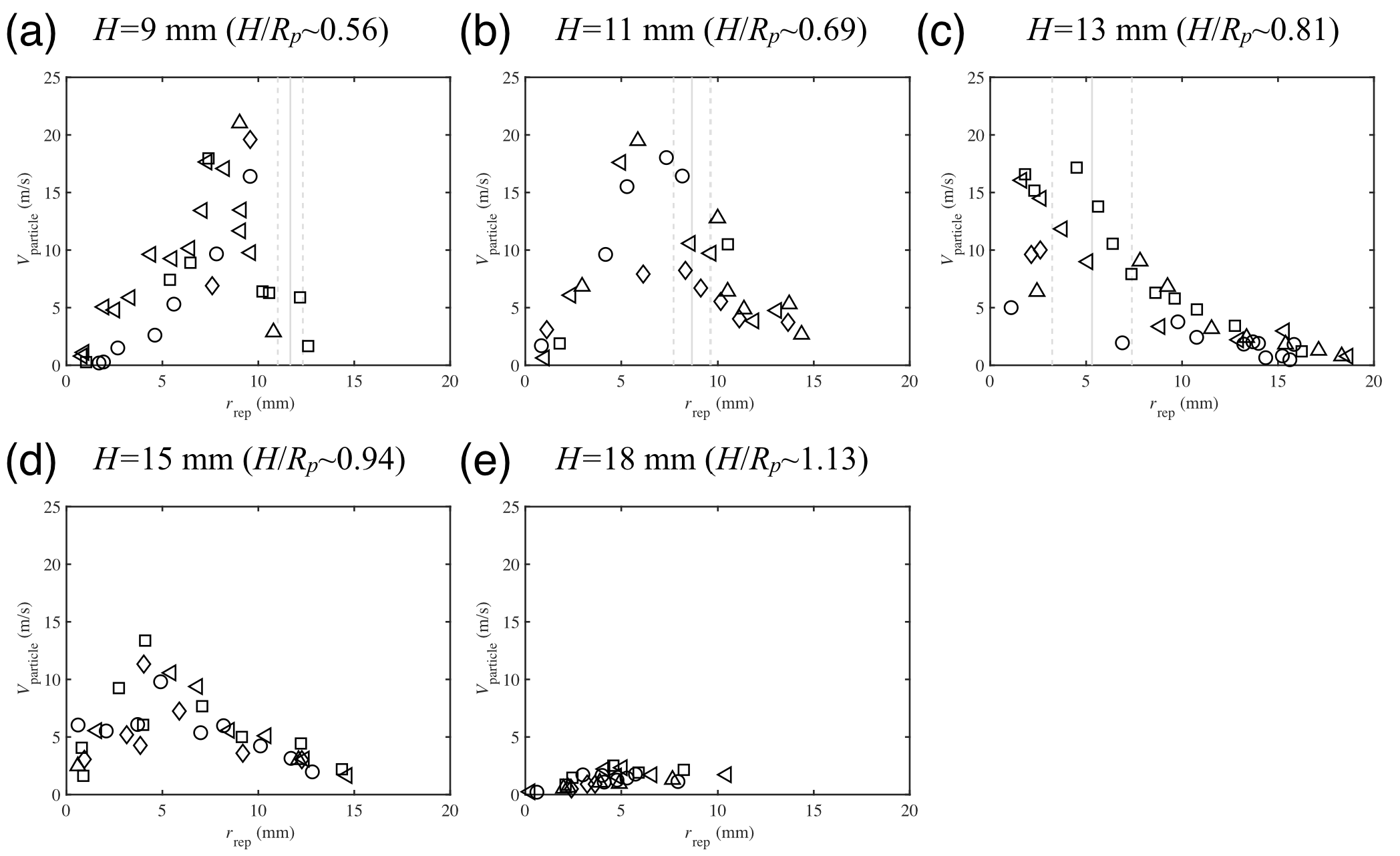}
    \caption{Particle speeds $V_\mathrm{particle}$ for various cases versus the characteristic particle location $r_\mathrm{rep}$. Each panel (a-e) contains the data from 5 different runs for each release height $H$. Solid and gashed gray lines (when available) represent the mean and standard deviation of $R_\mathrm{rim}$ data. We observed a formation of either toroidal cavitation or the transition cavitation upon the first stretch of the popper for (a-d) ($0.56\leq~H/R_p\leq0.94$), while a ring bubble was formed when $H/R_p\sim1.13$.}
    \label{fig:VparticlevsRrep}
\end{figure*}

As represented by figure \ref{fig:PTV}(e), the particle speed $V_\mathrm{particle}$ reaches the maximum at a certain distance $r$ from the popper center. It is interesting to note that the peak shifts to a smaller $r$ as $H$ increases (figures \ref{fig:VparticlevsRrep}(a-c)). This is intuitive from the trend shown in figure \ref{fig:velocity}(b), i.e., the radius of the thinnest gap $R_\mathrm{rim}$ shrinks as $H/R_p$ increases. Solid and dashed gray lines in figure \ref{fig:VparticlevsRrep}(a-c) represent the mean and standard deviation of $R_\mathrm{rim}$ estimated by the three-dimensional imaging (figure \ref{fig:velocity}(b)), showing a qualitative agreement. $R_\mathrm{rim}$ values for $H/R_p\sim0.94$ and 1.13 were not available as discussed. They showed the reduction in speed for larger $H/R_p$ and the outward shift of the peak.

\subsection{Force upon the popper snap}
\begin{figure*}
    \centering
\includegraphics[width=0.85\paperwidth]{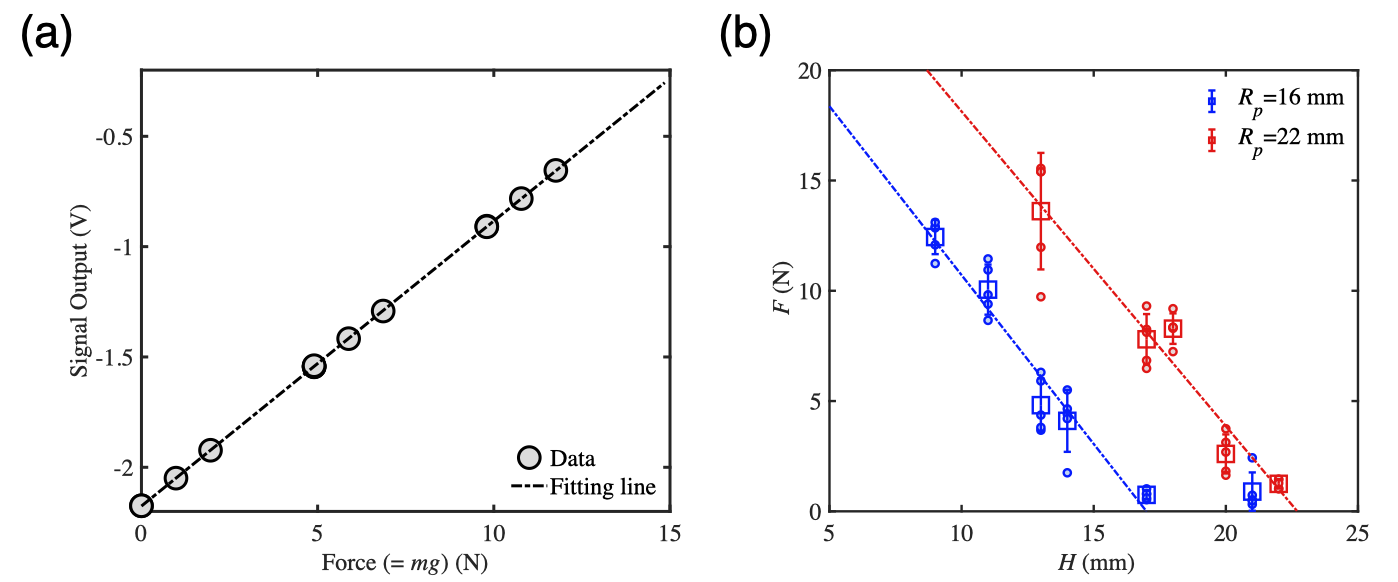}
    \caption{(a) A calibration curve between the signal output from the force sensor ($S.O.$) and the force estimated by the weight ($F=mg$). The fitting line is estimated to be $S.O.=0.1292F-2.1757$. (b) Measured forces as a function of the platform position $H$, where the colors distinguish the popper sizes. The trend lines are estimated to be $F=-1.531H+26.02$ ($R^2=0.9188$) for $R_p=16$~mm (blue) and $F=-1.427H+32.4$ ($R^2=0.8869$) for $R_p=22$~mm (red). We note that we excluded the data for $H=21$~mm when obtaining the fitting curve for a smaller popper ($R_p=16$~mm). We note that we have five trials at each condition (marked by dots) to obtain the average value (square) and standard deviation (error bar).}
    \label{fig:force}
\end{figure*}

Calibration was performed by using the weight balances to convert units from signal output ($S.O.$) (V) to force (N), where the result is shown in figure \ref{fig:force}(a). We varied the mass of the weight balances from 0.2 to 1.2 kg. The calibration curve (a dashed line) is $S.O.=0.1292F-2.1757$. The dimensional result based on the calibration curve is shown in figure \ref{fig:force}(b) for both popper sizes. Two dashed lines are demonstrating the downward trends. It was obvious that the impulsive force $F$ becomes smaller as $H$ increases. Also, the larger popper ($R_p=22$~mm) can cause a larger force when compared to the one with a smaller popper ($R_p=16$~mm). Figure \ref{fig:force}(b) demonstrates that the height $H$ and the popper size $R_p$ govern the impulsive force $F$ as discussed in the main manuscript. Note that, we assumed the scaling relationship between $F$ and $H/R_p$ does not change in either air or underwater, in order to apply the findings to the current problem.

%

\end{document}